\documentclass[12pt]{article}
 \usepackage{epsfig}
 \def\be{\begin{equation}}
 \def\ee{\end{equation}}
 \def\bea{\begin{eqnarray}}
 \def\eea{\end{eqnarray}}
 \usepackage{graphicx}

 \catcode`\@=11
 \def\lsim{\mathrel{\mathpalette\@versim<}}
 \def\gsim{\mathrel{\mathpalette\@versim>}}
 \def\@versim#1#2{\vcenter{\offinterlineskip
 \ialign{$\m@th#1\hfil##\hfil$\crcr#2\crcr\sim\crcr } }}
 \catcode`\@=12

 \catcode`@=12
\begin{document}
 \thispagestyle{empty}
 \begin{flushright}
 UCRHEP-T586\\
 Jan 2018\
 \end{flushright}
 \vspace{0.6in}
 \begin{center}
 {\LARGE \bf $[SU(2)]^3$ Dark Matter\\}
 \vspace{1.2in}
 {\bf Ernest Ma\\}
 \vspace{0.2in}
{\sl Physics and Astronomy Department,\\ 
University of California, Riverside, California 92521, USA\\}
\vspace{0.1in}
{\sl Jockey Club Institute for Advanced Study,\\ 
Hong Kong University of Science and Technology, Hong Kong, China\\} 
\end{center}
 \vspace{1.2in}

\begin{abstract}\
An extra $SU(2)_D$ gauge factor is added to the well-known left-right 
extension of the standard model (SM) of quarks and leptons.  Under 
$SU(2)_L \times SU(2)_R \times SU(2)_D$, two fermion bidoublets 
$(2,1,2)$ and $(1,2,2)$ are assumed.  The resulting model has an 
automatic dark $U(1)$ symmetry, in the same way that the SM has automatic 
baryon and lepton $U(1)$ symmetries.  Phenomenological implications are 
discussed, as well as the possible theoretical origin of this proposal.
\end{abstract}

 \newpage
 \baselineskip 24pt

\noindent \underline{\it Introduction}~:~
In the standard model (SM) of quarks and leptons, the choice of the gauge 
symmetry, i.e. $SU(3)_C \times SU(2)_L \times U(1)_Y$, and the particle 
content, i.e. quarks and leptons:
\begin{eqnarray}
&& (u,d)_L \sim (3,2,1/6), ~~~ u_R \sim (3,1,2/3), ~~~ d_R \sim (3,1,-1/3), 
\\
&& (\nu,l)_L \sim (1,2,-1/2), ~~~ l_R \sim (1,1,-1),
\end{eqnarray}
together with the one Higgs scalar doublet
\begin{equation}
\Phi = (\phi^+,\phi^0) \sim (1,2,1/2),
\end{equation}
automatically imply the existence of two global $U(1)$ symmetries, i.e. 
baryon number ($B$) under which quarks have charge $1/3$, and lepton 
number ($L$) under which leptons have charge $1$.  Is there a corresponding 
scenario for the existence of dark matter?  Consider for example the 
conventional left-right extension of the SM.  Because of the implied 
$U(1)_{B-L}$ gauge factor, a discrete $Z_2$ parity, i.e. 
$R = (-1)^{3B+L+2j}$, may be used to distinguish some new particles from 
those of the SM automatically.  The importance of this observation is 
that this parity is not imposed, as is necessary in the minimal 
supersymmetric standard model, or in models of dark matter~\cite{m15} 
assuming only the SM gauge symmetry.  Whereas this idea 
of an automatic $R$ parity has been implemented in some recent 
studies~\cite{hp15,gh16,ahr17,kmppz18,dhqvv18}, I look instead 
in this paper for a dark $U(1)$ symmetry (and not just a dark parity) 
which is also unrelated to $B$ or $L$, but on the same footing, i.e. 
its emergence as the result of gauge symmetry and particle content.   
In the following I show how it may be achieved by inserting an extra 
$SU(2)_D$ gauge factor to the well-known 
$SU(3)_C \times U(1)_{B-L} \times SU(2)_L \times SU(2)_R$ model.  Its 
theoretical origin is a possible $SU(6)$ generalization of the Pati-Salam 
$SU(4)$ symmetry~\cite{ps74}.

\noindent \underline{\it Particle Content}~:~
Under 
$SU(3)_C \times U(1)_{B-L} \times SU(2)_L \times SU(2)_R \times SU(2)_D$, 
the quarks and leptons transform as expected, i.e. as singlets under 
$SU(2)_D$:
\begin{eqnarray}
&& (u,d)_L \sim (3,1/6,2,1,1), ~~~ (u,d)_R \sim (3,1/6,1,2,1), \\ 
&& (\nu,l)_L \sim (1,-1/2,2,1,1), ~~~ (\nu,l)_R \sim (1,-1/2,1,2,1),
\end{eqnarray}
and the new fermions transform as bidoublets:
\begin{equation}
\pmatrix{\psi_1^0 & \psi_2^+ \cr \psi_1^- & \psi_2^0}_L \sim (1,0,2,1,2), ~~~ 
\pmatrix{\psi_3^0 & \psi_4^+ \cr \psi_3^- & \psi_4^0}_R \sim (1,0,1,2,2), 
\end{equation}
where $SU(2)_{L,R}$ act vertically, and $SU(2)_D$ horizontally.  The 
electric charge is given by
\begin{equation}
Q = {1 \over 2}(B-L) + I_{3L} + I_{3R} + I_{3D}.
\end{equation}
The gauge symmetry is broken by one $SU(2)_R$ doublet, and two 
$SU(2)_L \times SU(2)_R$ bidoublets:
\begin{equation}  
\pmatrix{\phi_R^+ \cr \phi_R^0} \sim (1,1/2,1,2,1), ~~~  
\pmatrix{\phi_1^0 & \phi_2^+ \cr \phi_1^- & \phi_2^0} \sim (1,0,2,2,1), ~~~ 
\pmatrix{\phi_3^0 & \phi_4^+ \cr \phi_3^- & \phi_4^0} \sim (1,0,2,2,1).
\end{equation}
Whereas the gauge $U(1)_{B-L}$ 
is broken, the global $U(1)$ symmetries of baryon number $(B)$ and lepton 
number $(L)$ remain. 

What about the extra fermion bidoublets?  The crucial observation is that 
they have built-in invariant masses because of the allowed terms
\begin{equation}
\psi^0_1 \psi^0_2 - \psi^-_1 \psi^+_2, ~~~ 
\psi^0_3 \psi^0_4 - \psi^-_3 \psi^+_4.
\end{equation}
At the same time, $\bar{\psi}_{1L} \psi_{3R}$ and 
$\bar{\psi}_{2L} \psi_{4R}$ acquire mass terms from the $\phi_{1,2,3,4}$ 
vacuum expectation values.  This means that an automatic global $U(1)_D$ 
symmetry emerges, i.e.
\begin{equation}
\psi_{1L}, ~ \psi_{3R} \sim -1, ~~~ \psi_{2L}, ~ \psi_{4R} \sim 1, 
\end{equation}
whereas all particles which are singlets under $SU(2)_D$ are trivial under it.  
It thus serves as a possible dark $U(1)$ symmetry unrelated to $B$ or $L$.  
The lighter of the two neutral Dirac fermion eigenstates is then 
a possible candidate for dark matter.  Since $\psi_{1,2}$ 
have $SU(2)_L$ interactions, they may scatter off nuclei with a large 
elastic cross section and are thus ruled out by direct-search 
experiments.  It is hence assumed that the dark matter is predominantly 
$\psi^0_{3,4}$.  At this stage, $SU(2)_D$ remains unbroken.  To break it, 
one $SU(2)_D$ Higgs triplet is added, i.e.
\begin{equation}
\pmatrix{\phi_D^{++} \cr \phi_D^+ \cr \phi_D^0} \sim (1,1,1,1,3).  
\end{equation}
This choice ensures that there is no coupling between $\Phi_D$ and the 
SM fermions, which would not be the case if it were a doublet.

\noindent \underline{\it Gauge Bosons}~:~
Masses of the gauge bosons come from the vacuum expectation values of the 
appropriate neutral scalar bosons.  Let
\begin{equation}
\langle \phi^0_{R,D,1,2,3,4} \rangle = v_{R,D,1,2,3,4}.
\end{equation}
The charged gauge bosons $W_D^\pm$ have mass $g_D^2v_D^2$ and 
does not mix with $W^\pm_{L,R}$, the $2 \times 2$ mass-squared matrix of 
which is given by
\begin{equation}
{\cal M}^2_{W_L - W_R} = \pmatrix {(1/2) g_L^2 (v_1^2 + v_2^2+v_3^2+v_4^2) & 
-g_L g_R (v_1 v_2+v_3v_4) \cr -g_L g_R (v_1 v_2 + v_3 v_4) & 
(1/2) g_R^2 (v_R^2 + v_1^2 + v_2^2+v_3^2+v_4^2)}.
\end{equation}
Since $W_D^+$ takes $\psi_{1,3}$ to $\psi_{2,4}$, it has charge $+2$ under 
$U(1)_D$ to conform with Eq.~(10) and $\phi_D^{++}$ has charge $+4$.  This 
shows that $U(1)_D$ is not broken by $\phi_D$.  Note that the mass 
degeneracy of $\psi^0_{3R}/\psi^0_{4R}$ 
with $\psi^-_{3R}/\psi^+_{4R}$ is broken by a small finite radiative 
correction~\cite{s95} through the exchange of neutral gauge bosons. 
Hence $\psi^-_{3R}$ decays to the invisible $\psi^0_{3R}$ and a virtual 
$W^-_R$ which may convert to $\bar{u} d$.  Its lifetime is presumably 
quite long and the outgoing lepton has rather low momentum because of 
the kinematics.  This kind of signature may be searched for at the 
Large Hadron Collider (LHC) as already pointed out~\cite{s95}.

There are four neutral gauge bosons, i.e. $B$ from $U(1)_{B-L}$, $W_{3L}$ from 
$SU(2)_L$, $W_{3R}$ from $SU(2)_R$, $W_{3D}$ from $SU(2)_D$, with couplings 
$g_B, g_L, g_R, g_D$ respectively.  Let them be rotated to the following four 
orthonormal states:
\begin{eqnarray}
A &=& {e \over g_B} B + {e \over g_L} W_{3L} + {e \over g_R} W_{3R} + 
{e \over g_D} W_{3D}, \\ 
Z &=& {e \over g_Y} W_{3L} - {e \over g_L} \left( {g_Y \over g_B} B + 
{g_Y \over g_R} W_{3R} + {g_Y \over g_D} W_{3D} \right), \\ 
Z_R &=& {g_R \over \sqrt{g_R^2+g_B^2}} W_{3R} - {g_B \over \sqrt{g_R^2+g_B^2}} 
B, \\ 
Z_D &=& \sqrt{1-{g_Y^2 \over g_D^2}} W_{3D} - {g_Y \over g_D} \left( 
{g_B \over \sqrt{g_R^2+g_B^2}} W_{3R} + {g_R \over \sqrt{g_R^2+g_B^2}} B \right),
\end{eqnarray}
where
\begin{eqnarray}
{1 \over e^2} = {1 \over g_L^2} + {1 \over g_Y^2},  
~~~ {1 \over g_Y^2} = {1 \over g_D^2} + {1 \over g_R^2} + {1 \over g_B^2}.
\end{eqnarray}
The mass terms are given by
\begin{eqnarray} 
&& {1 \over 2} (g_B B - g_R W_{3R})^2 v_R^2 + 2 (g_B B - g_D W_{3D})^2 v_D^2 
\nonumber \\ 
&& + ~{1 \over 2} (g_L W_{3L} - g_R W_{3R})^2 (v_1^2+ v_2^2+ v_3^2 + v_4^2). 
\end{eqnarray}
It is easily shown that the photon $A$ is massless and decouples from 
$Z,Z_R,Z_D$ as it should.  The $3 \times 3$ mass-squared matrix spanning 
the latter is given by
\begin{eqnarray}
 {\cal M}^2_{ZZ} &=& {1 \over 2} (g_L^2+g_Y^2)(v_1^2 + v_2^2 + v_3^2 + v_4^2), \\ 
 {\cal M}^2_{Z_R Z_R} &=& {1 \over 2} (g_R^2+g_B^2) v_R^2 + {4g_B^4 v_D^2 + 
g_R^4 (v_1^2+v_2^2 + v_3^2 + v_4^2) \over 2(g_R^2+g_B^2)}, \\ 
 {\cal M}^2_{Z_D Z_D} &=& {g_D^2 g_R^2 g_B^2 \over 2g_Y^2 (g_R^2+g_B^2)} (4v_D^2)
 + {g_Y^2 g_R^2 g_B^2 \over 2g_D^2 (g_R^2+g_B^2)} (v_1^2+v_2^2+v_3^2+v_4^2), \\ 
 {\cal M}^2_{ZZ_R} &=& -{g_L g_Y g_R^2\over 2e \sqrt{g_R^2+g_B^2}}  
(v_1^2+v_2^2+v_3^2+v_4^2), \\ 
 {\cal M}^2_{ZZ_D} &=& {e g_Y^2 g_R g_B \over 2 g_L g_D 
\sqrt{g_R^2+g_B^2}} (v_1^2 + v_2^2+v_3^2+v_4^2), \\ 
 {\cal M}^2_{Z_R Z_D} &=& {g_R g_B \over 2(g_R^2+g_B^2)} \left[ {g_D g_B^2 
\over g_Y} (4v_D^2) - {g_Y g_R^2\over g_D} (v_1^2+v_2^2+v_3^2+v_4^2) 
\right].
\end{eqnarray}
To ensure that $SU(2)_L$ is broken at a scale significantly lower than that 
of $SU(2)_R$ or $SU(2)_D$, it is assumed that
\begin{equation}
v_1^2+v_2^2+v_3^2+v_4^2 << v_R^2, ~ v_D^2.
\end{equation}
Hence $Z$ decouples effectively from $Z_R$ and $Z_D$, with negligible 
mixing to the latter.  In the remaining $Z_R-Z_D$ sector, if the 
$v_1^2+v_2^2+v_3^2+v_4^2$ terms are neglected, then the $2 \times 2$ 
mass-squared matrix is of the form
\begin{equation}
{\cal M}^2_{Z_R-Z_D} = \pmatrix{A+B & \sqrt{BC} \cr \sqrt{BC} & C},
\end{equation}
where
\begin{equation}
A = {1 \over 2}(g_R^2+g_B^2)v_R^2, ~~ B = {g_B^4 \over 2(g_R^2+g_B^2)} (4 v_D^2), 
~~ C = {g_R^2 g_D^2 \over g_Y^2 g_B^2} B.
\end{equation}
There are two interesting limits.
\begin{itemize}
\item{(1) $B,C << A$, then $A$ and $C$ are eigenvalues with $Z_R$ and $Z_D$ 
as eigenstates.}
\item{(2) $A << B,C$, then $B+C$ and $AC/(B+C)$ are eigenvalues with 
$Z_1 = (g_Yg_B Z_R + g_R g_D Z_D)/\sqrt{g_Y^2 g_B^2 + g_R^2 g_D^2}$ and 
$Z_2 = (g_Rg_D Z_R - g_Y g_B Z_D)/\sqrt{g_Y^2 g_B^2 + g_R^2 g_D^2}$ as 
eigenstates.}
\end{itemize}

\noindent \underline{\it Gauge Interactions}~:~
The neutral-current gauge interactions are given by
\begin{eqnarray}
{\cal L}_{NC} &=& e A j_{em} + g_Z Z (j_{3L} - \sin^2 \theta_W j_{em}) + 
{1 \over \sqrt{g_R^2+g_B^2}} Z_R (g_R^2 j_{3R} - g_B^2 j_B) \nonumber \\ 
&+& g_Y Z_D \left( {g_D \sqrt{g_R^2+g_B^2} \over g_R g_B} j_{3D} - 
{g_R g_B \over g_D \sqrt{g_R^2+g_B^2}} (j_{3R} + j_B) \right).
\end{eqnarray}
In particular $Z_2$ couples to 
\begin{equation}
{g_R \sqrt{g_Y^2 g_B^2 + g_R^2 g_D^2} \over g_D \sqrt{g_R^2+g_B^2}} j_{3R}   
- {g_Y^2 g_D \sqrt{g_R^2+g_B^2} \over g_R \sqrt{g_Y^2 g_B^2 + g_R^2 g_D^2}} 
(j_{3D}+j_{B}).
\end{equation}

If $v_D^2 << v_R^2$, then $Z_D$ is the much lighter mass eigenstate with mass 
given by Eq.~(22).  It couples to quarks and leptons according to Eq.~(29) 
with 
\begin{eqnarray} 
j_{3R} &=& {1 \over 2} \bar{u}_R \gamma u_R - {1 \over 2} \bar{d}_R 
\gamma d_R + {1 \over 2} \bar{\nu}_R \gamma \nu_R - {1 \over 2} 
\bar{l}_R \gamma l_R, \\
j_{B} &=& {1 \over 6} (\bar{u} \gamma u + \bar{d} \gamma d) 
- {1 \over 2} (\bar{\nu} \gamma \nu + \bar{l} \gamma l), \\ 
 j_{3D} &=& 0.
\end{eqnarray}
For the dark Dirac fermion $\psi_3/\psi_4$,
\begin{equation}
j_{3R} = -j_{3D} = {1 \over 2} \bar{\psi}_{3R} \gamma \psi_{3R} - {1 \over 2} 
\bar{\psi}_{4R} \gamma \psi_{4R}, ~~~ j_B=0.
\end{equation}
At the LHC, $Z_D$ may be observed through its 
production by $u$ and $d$ quarks, with its subsequent decay to lepton 
pairs.  The $c_{u,d}$ coefficients~\cite{ATLAS14,CMS15} used in the data 
analysis are 
\begin{eqnarray}
c_u &=& (g^2_{uL} + g^2_{uR})B = {g_Y^2 g_R^2 g_B^2 \over g_D^2 (g_R^2+g_B^2)} 
\left[ \left( {1 \over 6} \right)^2 + \left( {2 \over 3} \right)^2 
\right] B, \\ 
c_d &=& (g^2_{dL} + g^2_{dR})B = {g_Y^2 g_R^2 g_B^2 \over g_D^2 (g_R^2+g_B^2)} 
\left[ \left( {1 \over 6} \right)^2 + \left( -{1 \over 3} \right)^2 \right] B,
\end{eqnarray}
where $B$ is the $Z_D$ branching fraction to $e^-e^+$ and $\mu^-\mu^+$. 
To estimate $c_{u,d}$, let $g_D = g_R = g_L$, then
\begin{equation}
{e^2 \over g_B^2} = 1 - {3 e^2 \over g_L^2} = 1 - 3(0.23) = 0.31.
\end{equation}
Assuming that $Z_D$ decays to 3 copies of the dark fermions of Eq.~(6) 
in addition to all the quarks and leptons, $B$ is estimated to be about 0.07,  
and $c_u =1.8 \times 10^{-3}$, $c_d = 5.4 \times 10^{-4}$.  Based on the 
13 TeV LHC data from ATLAS~\cite{ATLAS17}, this translates to a bound of 
about 3.5 TeV on the $Z_D$ mass. 

If $v_R^2 << v_D^2$, then $Z_2$ is the much lighter mass eigenstate with mass 
given by
\begin{equation}
M^2_{Z_2} = {g_R^2 g_D^2 (g_R^2 + g_B^2) \over 2(g_Y^2 g_B^2 + g_R^2 g_D^2)} 
v_R^2 = 0.304~v_R^2.
\end{equation}
It couples to fermions according to Eq.~(30).  The branching fraction $B$ 
is then about 0.03, and the $c_{u,d}$ coefficients are $1.6 \times 10^{-3}$ 
and $2.8 \times 10^{-3}$ respectively.  This translates to a bound 
of about 3.6 TeV on the $Z_2$ mass.  Note that this bound depends on 
$c_u$ more than $c_d$ because the LHC is a proton collider.

\noindent \underline{\it Dark Matter Interactions}~:~
The particles beyond the conventional left-right model are the $SU(2)_D$ 
gauge bosons, the $\psi$ fermions and the one Higgs scalar $\Phi_D$ triplet.  
Whereas $SU(2)_D$ is completely broken by $\Phi_D$, a residual global 
$U(1)_D$ symmetry remains, under which
\begin{equation}
\psi_{1L},\psi_{3R} \sim -1, ~~~ \psi_{2L},\psi_{4R} \sim +1, ~~~ 
W_D^\pm \sim \pm 2, ~~~ \phi_D^{\pm \pm} \sim \pm 4, 
\end{equation}
and the neutral $W_{3D}$ and the physical neutral scalar $h_D$ are trivial, 
which allow them to mix with the other neutral gauge bosons and scalar 
bosons.  The dark Dirac fermion $\psi$ is assumed to be dominantly 
composed of $\psi_{3R}$ and $\psi_{4R}$.  To be specific, the outgoing 
$\psi_{4R}$ may be redefined as an incoming $\psi_{3L}$, in which case 
the Dirac fermion $\psi$ has a vector coupling to 
$g_R W_{3R} - g_D W_{3D}$.

The elastic scattering of $\psi$ off nuclei in underground direct-search 
experiments is possible through $Z_D$ or $Z_2$.  The spin-independent 
cross section $\sigma_0$ is enhanced by coherence and depends only on 
their vector couplings to the $u$ and $d$ quarks.  For $Z_D$ which couples to 
$0.547 j_{3D} - 0.233 (j_{3R}+j_B)$,
\begin{equation}
u_V = -0.0971, ~~~ d_V = 0.0194, ~~~ \psi_V = -0.390.
\end{equation}
For $Z_2$ which couples to $0.547 j_{3R} - 0.233 (j_{3D}+j_B)$,
\begin{equation}
u_V = 0.0979, ~~~ d_V = -0.1756, ~~~ \psi_V = 0.390.
\end{equation}
The cross section $\sigma_0$ is then given by 
\begin{equation}
\sigma_0 = {4 \mu^2 \over \pi A^2} [Z f_p + (A-Z)f_n]^2,
\end{equation}
where $\mu$ is the reduced mass of the effective interaction and equal to 
the nucleon mass for large $m_\psi$. 
In the case of $Z_D$ as the mediator,
\begin{equation}
f_p = {\psi_V (2u_V + d_V) \over M^2_{Z_{D}}} = {0.0682 \over M^2_{Z_D}}, ~~~ 
f_n = {\psi_V (u_V + 2d_V) \over M^2_{Z_{D}}} = {0.0227 \over M^2_{Z_D}}.
\end{equation}
In the case of $Z_2$ as the mediator,
\begin{equation}
f_p = {\psi_V (2u_V + d_V) \over M^2_{Z_{2}}} = {0.0079 \over M^2_{Z_2}}, ~~~ 
f_n = {\psi_V (u_V + 2d_V) \over M^2_{Z_{2}}} = -{0.0988 \over M^2_{Z_2}}.
\end{equation}
Assuming $m_\psi = 150$ GeV for example, $\sigma_0$ is bounded by the latest 
experimental result~\cite{x1t17} to be below $2 \times 10^{-46}$ cm$^2$. 
Using $Z=54$ and $A=131$ for xenon, this translates to $M_{Z_D} > 7.8$ TeV 
and $M_{Z_2} > 9.0$ TeV, which are 
stronger than the LHC bounds discussed earlier.

Instead of $Z_D$ or $Z_2$, if the lightest new neutral gauge boson is 
$Z_3 = (g_B g_D Z_R + g_Y g_R Z_D)/\sqrt{g_Y^2 g_R^2 + g_B^2 g_D^2}$, then 
it is easily shown from Eq.~(29) that it couples to
\begin{equation}
{g_Y^2 g_D \sqrt{g_R^2+g_B^2} \over g_R \sqrt{g_Y^2 g_R^2 + g_B^2 g_D^2}} 
(j_{3R}+j_{3D}) - {g_B \sqrt{g_Y^2 g_R^2 + g_B^2 g_D^2} \over g_D 
\sqrt{g_R^2+g_B^2}} j_{B}.
\end{equation}
This means that $\psi_V = 0$ and there would be no interaction through 
$Z_3$ with nuclei and no bound on the mass of $Z_3$ from direct-search 
experiments.  In other words, if the lightest new neutral gauge boson 
has a dominant $Z_3$ component, its bound may be lowered to a value 
comparable to that from the LHC.

Consider now the relic abundance of $\psi$.  Its annihilation cross section 
through any new neutral gauge boson is much below 1 pb for a gauge-boson 
mass greater than 3.5 TeV.  Hence a different process is required.  
Consider then the Yukawa sector.  Note first that there is no scalar 
singlet, so if the dark 
fermion $\psi$ is composed of only $\psi^0_{3R}$ and $\psi^0_{4R}$ with 
the invariant mass term $\psi^0_{3R} \psi^0_{4R}$, it has 
no $\bar{\psi} \psi$ coupling to any scalar.  However, as pointed out 
already, there are also the allowed 
$\bar{\psi}^0_{3R}(\bar{\phi}_1^0 \psi^0_{1L} + \phi_1^+ \psi^-_{1L})$ + 
$\bar{\psi}^0_{4R}(\bar{\phi}_2^0 \psi^0_{2L} + \phi_2^- \psi^+_{2L})$ 
and 
$\bar{\psi}^0_{3R}(\bar{\phi}_3^0 \psi^0_{1L} + \phi_3^+ \psi^-_{1L})$ + 
$\bar{\psi}^0_{4R}(\bar{\phi}_4^0 \psi^0_{2L} + \phi_4^- \psi^+_{2L})$ 
terms.  Hence $\psi$ annihilation to scalars is possible and it may remain 
in thermal equilibrium in the early Universe until the temperature drops 
below $m_\psi$.

There are several diagrams for 
$\psi$ annihilation to scalars.  As an estimate, consider Fig.~1 which 
depicts the process $\psi \bar{\psi} \to \phi^+ \phi^-$ through $\psi^-$ 
exchange.
\begin{figure}[htb]
\vspace*{-5cm}
\hspace*{-3cm}
\includegraphics[scale=1.0]{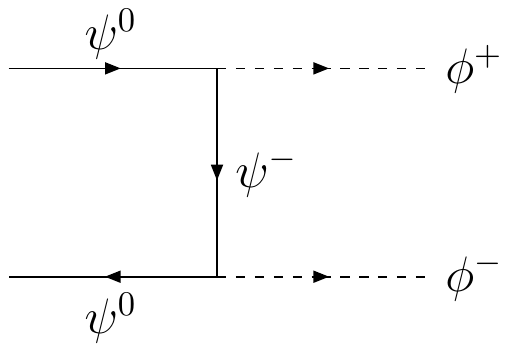}
\vspace*{-21.5cm}
\caption{Dark fermion annihilation to scalars.}
\end{figure}
The cross section $\times$ relative velocity is given by
\begin{equation}
\sigma v_{rel} = {f^4 \over 16 \pi} \left( 1 - {m_\phi^2 \over m_\psi^2} 
\right)^{3/2} {m_\psi^2 \over (M^2+m_\psi^2-m_\phi^2)^2},
\end{equation}
where $f$ is the $\bar{\psi}^0 \psi^- \phi^+$ coupling and $M$ is 
the mass of the exchanged $\psi^-$.  As an example, let $m_\psi = 150$ GeV, 
$m_\phi = 100$ GeV, and $M = 200$ GeV, then $\sigma v_{rel} = 1$ pb is 
obtained for $f=0.442$.  This shows that the proper relic abundance of 
dark matter in the Universe is possible within this framework.

\noindent \underline{\it Theoretical Origin of $SU(2)_D$}~:~
As presented, the introduction of $SU(2)_D$ and the new fermions of 
Eq.~(6) seems rather {\it ad hoc}.  However, there is a possible 
unifying theoretical framework underlying their existence.  Consider 
the well-known Pati-Salam partial unification symmetry 
$SU(4) \times SU(2)_L \times SU(2)_R$~\cite{ps74}, under which quarks and 
leptons are organized according to
\begin{eqnarray}
\pmatrix{u & u & u & \nu \cr d & d & d & l}_L \sim (4,2,1),  
~~~ \pmatrix{u & u & u & \nu \cr d & d & d & l}_R \sim (4,1,2),  
\end{eqnarray}
where $SU(4)$ contains $SU(3)_C \times U(1)_{B-L}$. 
If this is extended to $SU(6) \times SU(2)_L \times SU(2)_R$, the new 
fermions introduced are naturally included, i.e.
\begin{eqnarray}
\pmatrix{u & u & u & \nu & \psi^0_1 & \psi_2^+ \cr d & d & d & l & 
\psi_1^- & \psi_2^0}_L \sim (6,2,1), ~~~ 
\pmatrix{u & u & u & \nu & \psi_3^0 & \psi_4^+ \cr d & d & d & l & 
\psi_3^- & \psi_4^0}_R \sim (6,1,2).
\end{eqnarray}
This points to the possible unity of matter with dark matter, as 
discussed previously~\cite{kmppz18,b12,m13}.

The only other possible (and very intriguing) $SU(6)$ assignment is
\begin{eqnarray}
\pmatrix{u & u & u & \nu & x_1 & x_2 \cr d & d & d & l & 
y_1 & y_2}_L \sim (6,2,1), ~~~ 
\pmatrix{u & u & u & \nu & x_3 & x_4 \cr d & d & d & l & 
y_3 & y_4}_R \sim (6,1,2),
\end{eqnarray}
where $x_i$ and $y_i$ have charges $1/2$ and $-1/2$ respectively, and 
$SU(2)_D$ is unbroken.  This is a realization of an idea proposed many 
years ago~\cite{fl90}, where color $SU(3)_q$ for quarks is matched with 
a parallel color $SU(3)_l$ for leptons.  Whereas $SU(3)_q$ is unbroken, 
$SU(3)_l$ is broken to $SU(2)_l$, thereby confining only two components 
of the fundamental fermion triplet, leaving the third component free 
as the observed lepton.  This notion of leptonic color may be 
unified~\cite{bmw04} under $[SU(3)]^4$, with interesting 
predictions~\cite{kmppz17} for a future $e^-e^+$ collider.

Since $SU(4)$ is isomorphic to $SO(6)$ and $SU(2) \times SU(2)$ is 
isomorphic to $SO(4)$, it is well-known that 
$SU(4) \times SU(2)_L \times SU(2)_R$ may be embedded into $SO(10)$. 
As for $SU(6) \times SU(2)_L \times SU(2)_R$, it is not clear which  
simple group may be a possible unification symmetry.  It must of 
course be at least rank 7.

\noindent \underline{\it Concluding Remarks}~:~
The notion is put forward that dark matter is intimately related to 
matter and the global $U(1)$ symmetry which allows it to be stable is an 
automatic consequence of gauge symmetry and particle content in the 
same way that baryon and lepton numbers are so in the standard model. 
A specific proposal is the addition of an $SU(2)_D$ gauge symmetry 
with new fermions which are bidoublets under $SU(2)_{L,R} \times SU(2)_D$. 
It is shown that with the complete breaking of the $SU(2)_D$ gauge symmetry 
by an $SU(2)_D \times U(1)_{B-L}$ scalar triplet, a global $U(1)_D$ symmetry 
remains for the new particles.  Dark matter thus emerges naturally within 
this framework.  Its phenomenology is discussed, as well as the intriguing 
possibility that it may have a theoretical origin in 
$SU(6) \times SU(2)_L \times SU(2)_R$, where $SU(6)$ is a generalization 
of the well-known Pati-Salam $SU(4)$ which unifies quarks and leptons.

\noindent \underline{\it Acknowledgement}~:~
This work was supported in part by the U.~S.~Department of Energy Grant 
No. DE-SC0008541.

\bibliographystyle{unsrt}

\end{document}